\documentclass[manuscript,nonacm]{acmart}
\usepackage{tabularx}
\usepackage{multirow}
\usepackage{threeparttable}

\usepackage{placeins}
\providecommand{\correspondingauthor}{}
\AtBeginDocument{%
  }

\begin{document}

\title{When Does GenAI Use Support Critical Thinking? A Cross-Cultural Structural Account of Transactive Memory with GenAI}

\author{Mahir Akgun}
\authornote{Both authors contributed equally to this research.}
\email{makgun@psu.edu}
\orcid{0000-0001-6884-0119}
\correspondingauthor
\authornotemark[1]
\affiliation{%
  \institution{The Pennsylvania State University}
  \city{University Park}
  \state{PA}
  \country{USA}
}

\author{Sacip Toker}
\affiliation{%
  \institution{Çankaya University }
  \city{Ankara}
  \country{Türkiye}}
\email{saciptoker@cankaya.edu.tr}

\renewcommand{\shortauthors}{Akgun and Toker}.

\begin{abstract}
Whether generative AI (GenAI) use helps or hinders critical thinking remains contested, and existing accounts emphasize exposure to the tool rather than how a person structures their engagement with it. Extending Transactive Memory Systems (TMS) theory to human-AI partnerships, we test a structural account of critical thinking in GenAI-assisted academic work among 891 college students in the United States and Türkiye. Self-appraised thinking ability and confidence in the partnership shaped how much practical experience a person built up with GenAI, but neither predicted critical thinking directly. That experience mattered only through two downstream behaviors, dividing and coordinating cognitive labor with the tool, and actively reworking its output rather than accepting it as given. A multigroup comparison found this structure largely invariant, with two theoretically interpretable exceptions. This study offers a structural, TMS-based account of critical thinking in GenAI-assisted work.
\end{abstract}




\maketitle

\section{Introduction}

Generative AI (GenAI) tools have moved quickly from novelty to routine infrastructure in knowledge work, reshaping how people perform tasks and create value, with the promise of widespread organizational and academic adoption. This shift is not simply a matter of AI doing more work. Knowledge workers increasingly move from task execution to an oversight role that centers on verifying and editing what the AI produces, rather than producing the work themselves. Whether this role helps or hinders people's capacity to think critically about their own work is a question the literature does not agree on. Gerlich \cite{Gerlich2025} documents a negative association between frequent AI tool use and critical thinking ability, mediated by cognitive offloading, while Suriano et al. \cite{Suriano2024} and Goyal \cite{Goyal2025} find that GenAI use can support critical thinking when engagement remains active and reflective, and Simkute et al. \cite{Simkute2024} locate the risk not in GenAI use itself but in how the human-AI interaction is designed.

This divergence suggests that GenAI's effect on critical thinking may depend less on exposure to the tool than on how a person structures their engagement with it. Transactive Memory Systems (TMS) theory, which originates in Wegner's \cite{Wegner1987} account of how partners divide and coordinate cognitive labor, offers one account of this structure when extended from human-human teams to human-AI partnerships. Confidence and self-efficacy in working with the tool relate to critical thinking enactment in ways that depend on what the confidence is directed at, sometimes supporting it and sometimes substituting for it \cite{Lee2025}. The degree to which a person's engagement with AI-generated output is anchored on its initial suggestion, versus actively reflected on and transformed, offers a further, largely separate account \cite{Doshi2024,Catena2026}. How these pieces jointly shape whether a person enacts critical thinking in GenAI-assisted work remains underexamined.

This literature is also overwhelmingly one-sided, and cross-cultural comparisons are rare. Whether these dynamics hold consistently across cultural contexts, for instance a higher-power-distance, more collectivist setting compared with a Western, individualist one, is largely untested. This gap mirrors a longer-documented pattern in psychology more broadly. Research samples have historically skewed toward Western, Educated, Industrialized, Rich, and Democratic (WEIRD) populations \cite{Henrich2010}. Nielsen et al. \cite{Nielsen2017} find that despite repeated calls for change, high-impact developmental psychology journals remain heavily reliant on this narrow population, with little evidence of meaningful movement toward more diverse samples. Kupferschmidt \cite{Kupferschmidt2019} reports that this imbalance has persisted even as the field has become more aware of it, leaving open how well many of psychology's findings extend beyond that narrow population. This pattern extends to human-computer interaction research itself. Linxen et al. \cite{Linxen2021}found that 73\% of CHI publications between 2016 and 2020 relied on Western participant samples, representing less than 12\% of the world's population.

The present study addresses both gaps. Building on a transactive-memory-based account of human-GenAI partnership, we test whether confidence and self-efficacy, the resulting partnership with the tool, and a person's reflective engagement with its output together explain Critical Thinking Enactment in GenAI-assisted academic work. Using structural equation modeling, we test this account across two national samples, a Western, individualist context (the United States) and a higher-power-distance, more collectivist one (Türkiye). This lets us examine whether the same account holds consistently across both contexts. In doing so, this study contributes a structural, TMS-based account of when and why critical thinking is enacted in GenAI-assisted academic work, in place of an exposure-based one, and offers one of the few tests of whether such an account generalizes across national cultural contexts.

\section{Relevant Studies}

\subsection{Critical Thinking Enactment in GenAI-Assisted Work}

Whether generative AI (GenAI) tools help or hinder critical thinking during knowledge work remains an open empirical question, with recent studies pointing in different directions depending on how AI is used. On one side, Gerlich \cite{Gerlich2025} documents a negative association between frequent AI tool use and critical thinking ability, mediated by cognitive offloading (i.e., the tendency to delegate cognitive effort to an external tool rather than exercise it oneself). On the other side, Suriano et al. \cite{Suriano2024} find that structured interaction with ChatGPT can promote complex critical thinking skills among students, and Goyal \cite{Goyal2025} frames AI as a potential "cognitive partner," with a systematic review linking AI use to gains in self-reflection — but only when the human retains an active, reflective role in the exchange. Simkute et al. \cite{Simkute2024} offer a complementary account rooted in decades of Human Factors research on automation. Introducing GenAI into a workflow shifts users from actively producing work to passively evaluating AI-generated output. They locate the source of this cost not in GenAI use itself but in how the interaction is designed, pointing to insufficient feedback, disrupted task sequences, and poorly allocated division of labor between user and system.

Taken together, this literature suggests that mere exposure to GenAI is not what determines whether critical thinking is enacted or eroded; what appears to matter is the structure of engagement (i.e., how a person organizes their reliance on, and evaluation of, the tool). This reframes the central question addressed by the present study: rather than asking whether GenAI use helps or harms critical thinking, we ask what structural and dispositional factors shape whether a person enacts critical thinking — operationalized here as actively evaluating and assessing the quality and reliability of GenAI output — during a GenAI-assisted task. Transactive Memory Systems (TMS) theory, which describes how people structure the division and coordination of cognitive labor within a partnership, offers a natural framework for this question when the partner is not another person but a GenAI tool.

\subsection{Transactive Memory Systems, from Human-Human to Human-AI}

Transactive Memory Systems (TMS) theory originates in Wegner's \cite{Wegner1987} account of how closely collaborating dyads develop a shared, relationally emergent system for encoding, storing, and retrieving knowledge that exceeds what either partner holds individually, in which each partner comes to know not all of the information itself but where within the partnership it is located. Lewis \cite{Lewis2003} operationalized this theoretical structure for empirical measurement, decomposing TMS into three interrelated dimensions: specialization (differentiated, non-redundant expertise across partners), coordination (the ability to enact that specialized knowledge smoothly, without excessive prompting or redundant effort), and credibility (each partner's confidence in the reliability of the other's contributions). These three dimensions have since been treated as jointly necessary, mutually reinforcing components of a functioning transactive memory system.

A growing body of work extends this framework from human-human teams to human-AI partnerships. Bienefeld et al. \cite{Bienefeld2023}, studying ICU teams with access to an AI diagnostic agent, found that teams generated more novel hypotheses and spoke up more often when the AI was integrated as a knowledge source, but only in already higher-performing teams. They attribute this pattern to the AI's capacity to hold knowledge beyond human reach and to whether team members understand how that knowledge is generated and can calibrate appropriate trust in it. Hopf et al. \cite{Hopf2024} formalize this extension with their Transactive Intelligent Memory Systems (TIMS) framework, arguing that humans incorporate an intelligent agent into their transactive memory when the agent demonstrates sufficient mastery of a task, particularly as that task's complexity and knowledge intensity increase. Gupta and colleagues' COHUMAIN research agenda \cite{Gupta2023} similarly theorizes AI's role within collective memory, attention, and reasoning processes, treating human-AI collaboration as a distributed cognitive system rather than a simple tool-use relationship.

Akgun and Toker \cite{Akgun2026} developed and cross-culturally validated the Transactive Memory System with GenAI (TMS-GenAI) instrument to capture this human-AI extension, operationalizing perceived ability to think and ability to remember as balanced cognitive dimensions of the partnership alongside Lewis's \cite{Lewis2003} specialization, coordination, and credibility dimensions and a measure of GenAI familiarity. Within this framework, TMS offers a mechanistic account of why GenAI's effect on critical thinking should depend on structure rather than exposure. Recognizing a GenAI tool as a credible knowledge partner is a precondition for actively evaluating its output rather than passively accepting it; specialization and coordination describe the behavioral division of labor through which that evaluation is enacted; and familiarity with the tool reflects the accumulated interaction history through which this structure develops. This account, however, leaves two questions open — what motivates a person to engage a GenAI tool as a knowledge partner in the first place, and what governs their orientation toward its output once engaged — that the following two subsections address in turn.

\subsection{Confidence and Self-Efficacy in AI-Assisted Work}

Lee et al. \cite{Lee2025} provide the most directly relevant empirical precedent for the role of confidence in this process. In a survey of 319 knowledge workers describing 936 concrete instances of GenAI-assisted task completion, they measured three distinct confidence referents — confidence in one's own ability to complete the task without AI assistance, confidence in the AI's capability to perform the task, and confidence in one's own ability to evaluate the AI's output. They found that these three facets relate to self-reported critical thinking enactment in different, even opposing, directions: confidence in the AI's own capability negatively predicted critical thinking, while self-confidence and confidence in evaluating the AI's output both positively predicted it; general trust in GenAI was not a significant predictor. This pattern suggests that confidence is not a uniformly protective or uniformly risky factor for critical engagement. Its effect depends on what, specifically, the confidence is directed at. Confidence in oneself and one's ability to judge appears to support active evaluation, whereas confidence placed in the tool itself appears to substitute for it.

This confidence-based account sits within a broader literature on self-efficacy in AI-assisted work. Wang and Chuang \cite{Wang2023} extend Bandura's self-efficacy tradition to AI-assisted performance with their AI Self-Efficacy Scale (AISE), defining it as individuals' belief in their ability to use and interact with AI. Consistent with social cognitive theory, they treat this belief as changeable rather than fixed, shaped by personal, behavioral, and social or environmental factors. Akben and Dong \cite{Akben2025}, across two studies, find that job complexity has a curvilinear, inverted-U relationship with AI self-efficacy, which peaks at moderate complexity and declines at both extremes, while AI self-efficacy itself positively predicts both AI adoption readiness and AI adoption behavior. Together, this literature motivates treating confidence as a distinct, multi-referent construct whose relationship to critical thinking enactment is not fixed in direction but depends on which referent — self, tool, or judgment of the tool — is engaged in a given interaction, and on the broader task conditions under which that confidence operates.

\subsection{Cognitive Autonomy and Reflective Engagement with AI Output}

A second, complementary line of inquiry concerns how a person orients toward GenAI-generated content once it is produced: whether their engagement is oriented around the AI's initial output as the anchor for their own thinking, or around actively transforming and going beyond it. Doshi and Hauser \cite{Doshi2024} conducted a preregistered online experiment on short-story writing in which some writers received AI-generated story ideas. Access to these ideas increased individual creativity, but it also measurably pulled writers' output toward the AI's own suggestions, an effect the authors term anchoring on generative AI ideas. They describe this pattern as a social dilemma in which individual writers benefit while the collective diversity of novel content produced across writers narrows. Barsky-Moore \cite{BarskyMoore2025} addresses a related concern in the context of AI-assisted research, discussing how anchoring bias can be mitigated through deliberate strategies such as actively questioning and reassessing AI's initial output and maintaining awareness of one's own assumptions and biases while evaluating it.

These two accounts map onto two theoretically distinct facets of engagement with AI-generated drafts: (1) an anchoring orientation, in which the AI's initial suggestion substantially shapes the direction and constrains the originality of the final work, and (2) a reflective-transformation orientation, in which the person actively considers alternatives beyond the AI's suggestion, substantively reworks it, and remains aware of the AI's influence while doing so. Catena's \cite{Catena2026} review of AI and human autonomy offers a related framework for this distinction. AI's mediating influence on a person's deliberation and decisions can affect conditions of autonomy such as competency (bypassing or improving one's own reasoning and deliberative skills) and independence (exerting undue or supportive external influence). An anchoring orientation is what results when that mediating influence goes unchecked, that is, when a person does not exercise their own reasoning in responding to the AI's output. A reflective-transformation orientation is a related but conceptually distinct response to that same influence. The person actively exercises their own reasoning to transform the AI's output into something substantively different, rather than leaving it as an unexamined AI product. This distinction directly relates to Critical Thinking Enactment.

Because Critical Thinking Enactment, as operationalized in the present study, centers on actively evaluating and assessing the quality and reliability of AI output, it shares clear conceptual ground with the reflective-transformation orientation described above — both describe the exercise of reasoned, deliberate judgment on the AI's output, rather than passive acceptance of it. Suriano et al.'s \cite{Suriano2024} finding, described above, that structured interaction promotes complex critical thinking specifically when reflective engagement is retained, reinforces this connection. An orientation toward the AI's output that goes beyond simply anchoring on the initial draft appears to be what allows critical evaluation to occur in the first place, rather than exposure or familiarity with the tool being sufficient on its own.

\subsection{Synthesis of the Relevant Studies}
Taken together, this literature suggests that the effect of GenAI on critical thinking (Section 2.1) can be resolved by looking beneath aggregate exposure to the structure of engagement itself. Transactive Memory Systems theory (Section 2.2) provides the organizing framework. Recognizing GenAI as a credible knowledge partner, and dividing and coordinating cognitive labor with it, is a precondition for active evaluation rather than passive acceptance of its output. Confidence (Section 2.3) offers one entry point into this structure — but a multi-referent one, whose relationship to critical thinking enactment depends on whether it is directed at oneself, at the tool, or at one's own capacity to judge the tool's output. Cognitive autonomy and reflective engagement (Section 2.4) offer a second, complementary entry point, describing whether a person's orientation toward the AI's output is anchored on its initial suggestion or actively transformed through reasoned judgment. Where a user brings sufficient confidence and reflective orientation to a well-coordinated human-AI partnership, GenAI use appears more likely to support critical thinking; where these structural and motivational conditions are absent, the same tool use risks the disengagement and cognitive outsourcing documented in the literature (e.g, \cite{Jose2025}). Whether this account, developed primarily from Western, largely US-based samples, generalizes across cultural contexts is the question addressed next.

\subsection{Cross-Cultural Considerations}

Whether the account developed in the preceding sections generalizes beyond a single cultural context is itself an empirical question with direct precedent in the TMS literature. Bachrach et al.'s \cite{Bachrach2019} meta-analysis of 76 studies finds that the strength of the TMS-to-performance relationship is not culturally invariant. It is stronger in national contexts with higher power distance and higher in-group collectivism. Jarvenpaa and Keating's \cite{Jarvenpaa2011} longitudinal case study of a U.S.-Romanian engineering team offers convergent, mechanism-level evidence for why: cultural differences in values, practices, and institutions had a major impact on TMS indicators of specialization, coordination, and credibility.

This cultural patterning is not unique to TMS. Ashraf et al. \cite{Ashraf2014} tested the Technology Acceptance Model across a high-power-distance, high-collectivism culture (Pakistan) and a low-power-distance, individualist culture (Canada), expecting the model's predictive power to weaken in the culturally distant sample. Contrary to that expectation, the model's overall predictive power held in both countries, but perceived usefulness (PU) and perceived ease of use (PEOU) traded roles across the two samples: PEOU predicted intention directly only in Pakistan, while PU predicted intention directly only in Canada. Together, this literature suggests that cultural context is more likely to reshape which specific relationships in a model hold, and how strongly, than to invalidate the model's structure completely.

\section{Present Study}

Building on the studies reviewed above, the present study asks two questions.

\begin{itemize}
    \item RQ1: How do confidence in working with a GenAI tool, cognitive autonomy, reflective engagement with AI's output, and the transactive-memory-based partnership a person develops with it relate to Critical Thinking Enactment in GenAI-assisted academic work?
    \item RQ2: Does this account of Critical Thinking Enactment hold consistently across cultural contexts?
The Method section that follows details how each construct is measured and the analytic strategy used to address these questions.
\end{itemize}

\section{Method}

\subsection{Research Design}

A correlational survey research design was used, whereby naturally occurring associations among two or more measured variables are examined without manipulating any of them \cite{Fraenkel2022}. A survey approach was selected because it permits efficient collection of self-reported, quantitative data on attitudes, perceptions, and behaviors from a sample large enough to support latent-variable and correlational analyses; consistent with a cross-sectional structure, data were collected at a single point in time rather than across repeated occasions, capturing a snapshot of associations without inferring causal or temporal precedence \cite{Creswell2018}. The purpose of this design was to reveal the cross-sectional associations among validated variables, consistent with the descriptive-correlational tradition in educational and social science research. 

\subsection{Data Collection}

Most constructs were operationalized using items from validated instruments in the Transactive Memory Systems (TMS-GenAI) and GenAI adoption literature. The exception is Cognitive Autonomy in GenAI-assisted Work (CA-GenAI), which was developed for the present study; its two-factor structure (GenAI Anchoring and Reflective Transformation) was established through exploratory factor analysis rather than drawn from a prior validated instrument (see Measures). The outcome variable, Critical Thinking Enactment, was treated as a binary indicator, which required the use of a robust categorical estimator. An online survey tool was used for data collection. The responses not fully completed were removed.

\subsection{Participants}

Data were collected from 917 respondents drawn from two national contexts: the United States and Türkiye. After excluding 26 respondents who did not meet the Critical Thinking Enactment exclusion rule described in Measures, the final analytic sample consisted of 891 respondents: 645 from the United States and 246 from Türkiye. The participants were traditional college students. A convenience sampling strategy was applied because the survey was administered online, which offers a practical, low-cost way to reach a large, geographically dispersed sample \cite{Jager2017}. Ethical approvals were obtained from both institutions, and informed consent was collected from participants. 

\subsection{Measures}

\noindent{\textbf{Critical Thinking Enactment:}} Critical thinking enactment was measured using a single dichotomous item: "Have you ever engaged in reflective or critical thinking (e.g., evaluating how you used a GenAI tool and assessing the quality and reliability of its outputs) when working on this academic task with GenAI tool?" Respondents who answered Yes were then presented with a follow-up multi-select item asking which reflective or critical-thinking tactic(s) they had used: "What type of reflective/critical thinking tactic(s) did you use for this academic task with GenAI? (Select all that apply)." The six response options described distinct tactics: (1) reflecting on facts or basic concepts and cross-checking AI output with other sources; (2) reflecting on the organization, summarization, and generalization of AI-generated content; (3) reflecting on how the AI applied high-level concepts and how they themselves applied that knowledge; (4) reflecting on the coherence and logical relationships among individual elements of the AI's output; (5) reflecting on how ideas were combined to form new meaning; and (6) reflecting on the overall quality of the work relative to academic standards and expectations, alongside an open-ended "Other" option. Respondents who selected none of these six options, and whose "Other" response (if provided) did not give clear evidence of genuine critical-thinking engagement, were excluded from the dataset under this exclusion rule. This exclusion rule was applied so that a Yes response to the preceding item would not be accepted at face value without corroborating evidence of the reflective process it was intended to capture.

This item served as the primary outcome variable in the structural model. Because this variable is binary, it was treated as a categorical endogenous indicator in the structural equation model and estimated accordingly (see Data Analysis), rather than as a continuous or multi-item latent construct. 
\vspace{3mm}

\noindent{\textbf{GenAI Collaboration Confidence:}} The questionnaire contains three items (see Appendix~A) suggested and used by \citet{Lee2025}, with a scale from \textit{Not at all confident} (1) to \textit{Extremely confident} (5). Originally, the questionnaire was developed as separate items to measure confidence level. However, in the present study, a single-factor solution for the three Confidence items was extracted using maximum likelihood estimation ($N = 891$). The data were minimally suitable for factor analysis, Bartlett's $\chi^2(3) = 318$, $p < .001$, KMO $= .618$. The single factor accounted for loadings ranging from $.465$ to $.768$. These results support a coherent, unidimensional GenAI Collaboration Confidence construct. Cronbach's $\alpha$ and McDonald's $\omega$ were $.615$ and $.632$, respectively. Although below the conventional $.70$ threshold, values in the $.60$--$.70$ range are considered acceptable, particularly in exploratory research and for short, three-item scales \citep{Griethuijsen2014,Taber2017}.

This single-factor solution combines three distinct confidence referents---self-efficacy without AI (Item 1), trust in AI's capability (Item 2), and evaluative/verification confidence in judging AI's output (Item 3). Although these items target different objects of confidence (self vs.\ tool vs.\ judgment-of-tool), their convergence onto a single factor (Item~1 $= .768$, Item~2 $= .560$, Item~3 $= .465$) indicates that respondents do not compartmentalize these referents. Unlike \citet{Lee2025}, who asked respondents to rate these referents with respect to specific, concrete instances of GenAI-assisted task completion, the present study asked respondents to rate their general tendency toward confidence in GenAI collaboration rather than confidence tied to any one task instance. This difference in framing is a reason the three referents converge here. When confidence is rated as a general tendency rather than judged against a specific task instance, respondents are less likely to distinguish between self, tool, and judgment-of-tool confidence, and more likely to report all three from a single, general sense of capability in working with GenAI. Combined with the small item pool, treating them as a single composite is the more conservative interpretation the data support.
\vspace{3mm}

\noindent{\textbf{Cognitive Autonomy in GenAI-assisted Work (CA-GenAI):}} There are six items in the questionnaire (see Appendix~A) with the scale from `Not at all (1)' to `To a great extent (5)'. The purpose of these items is to gauge the extent to which participants feel that using GenAI shapes the direction of their final work, and the extent to which they actively reflect on and go beyond the AI-generated content in producing it. A two-factor solution for the six items was extracted using principal axis factoring with oblimin rotation ($N = 891$). The data were suitable for factor analysis, Bartlett's $\chi^2(15) = 1054$, $p < .001$, KMO $= .732$. The two factors accounted for $42.9\%$ of total variance (Factor~1 $= 22.8\%$, Factor~2 $= 20.1\%$) and showed a clean simple structure with no meaningful cross-loadings.

This two-factor structure aligns with the anchoring/reflective-transformation distinction developed in the literature review (Section~2.4; \citealp{Doshi2024,BarskyMoore2025}). GenAI Anchoring (see Appendix~A) captures the degree to which the AI draft shapes direction, refines rather than explores alternatives, and limits originality; Reflective Transformation (see Appendix~A) captures considering alternatives beyond the AI's suggestion, substantively transforming it, and remaining aware of that influence.

Factor~1, labeled GenAI Anchoring (loadings $\lambda = .422$--$.773$), while Factor~2, labeled Reflective Transformation (loadings $\lambda = .549$--$.668$). The inter-factor correlation was moderate ($\varphi = .491$). This value, obtained from principal axis factoring with direct oblimin rotation, is a model-implied parameter that reflects the relationship between the ``true'' (theoretical) common factors, derived directly from the rotation solution as it optimized the pattern matrix \citep{Grice2001}. This finding is consistent with the two factors representing distinct but co-occurring facets of engagement with GenAI-generated drafts, rather than opposing poles. This pattern suggests that GenAI Anchoring and Reflective Transformation are not opposite ends of one behavior. Instead, both appear to reflect how intensely a person engages with the AI's draft in the first place. Respondents who were more strongly anchored by the AI's draft also tended to transform it more actively and stay more aware of its influence, rather than less. This supports treating GenAI Anchoring and Reflective Transformation as correlated-but-distinct facets of engagement, with Reflective Transformation aligning conceptually with critical-thinking constructs like reflective judgment.

The overall measure showed satisfactory reliability (Cronbach's $\alpha = .715$, McDonald's $\omega = .720$). Reliability was lower for the two three-item subfactors (GenAI Anchoring: $\alpha = .668$, $\omega = .692$; Reflective Transformation: $\alpha = .667$, $\omega = .671$), consistent with the same short-scale pattern noted above for the Confidence measure. Alpha in particular tends to underestimate reliability as the number of items decreases \citep{Cortina1993,Loewenthal2020}, and few-item constructs are considered acceptable at an early stage of scale validation when the construct is narrow, well-understood, and theoretically grounded \citep{Diamantopoulos2012}---as is the case here, given the clean two-factor structure with no meaningful cross-loadings reported above.
\vspace{3mm}

\noindent{\textbf{Transactive Memory System with GenAI (TMS-GenAI): }} There are 27 items with the scale: Strongly disagree (1), Disagree (2), Neutral (3), Agree (4), Strongly agree (5). A six-factor solution for the 27-item pool was extracted using principal axis factoring with oblimin rotation ($N = 891$). The data were suitable for factor analysis, Bartlett's $\chi^2(351) = 10956$, $p < .001$, KMO $= .906$. The six factors jointly accounted for $51.7\%$ of total variance and were labeled Ability to Think ($\lambda = .602$--$.787$), Ability to Remember ($\lambda = .597$--$.914$), GenAI Familiarity ($\lambda = .638$--$.822$), Credibility ($\lambda = .583$--$.799$), Specialization and Coordination ($\lambda = .331$--$.537$), and Collaboration ($\lambda = .394$--$.507$). Two items were excluded due to weak loadings. These results support a coherent, six-dimensional structure spanning individual self-efficacy and human-AI transactive memory constructs, consistent with \citet{Wegner1987}'s original conceptualization of transactive memory as a relationally emergent structure.

The overall measure showed satisfactory reliability, with Cronbach's $\alpha$ and McDonald's $\omega$ of $.898$ and $.903$, respectively. Reliability was similarly satisfactory across all six sub-factors: Ability to Think ($\alpha = .867$, $\omega = .870$), Ability to Remember ($\alpha = .900$, $\omega = .904$), GenAI Familiarity ($\alpha = .847$, $\omega = .848$), Credibility ($\alpha = .773$, $\omega = .775$), Collaboration ($\alpha = .725$, $\omega = .736$), and Specialization and Coordination ($\alpha = .708$, $\omega = .720$).

\subsection{Constructing Retention and Model Specification Procedure}

Our initial data collection was informed by a broad, theoretically motivated set of constructs spanning transactive memory systems, GenAI-specific competence and confidence, and reflective/critical engagement with AI-generated content (see Section 2). To arrive at a single, testable structural model from this larger pool, we applied a three-step procedure, described below.

\subsubsection{Step 1 — Bivariate Association with Critical Thinking Enactment.}

Each candidate construct was tested for a statistically significant zero-order correlation with Critical Thinking Enactment, the study's focal outcome. This step follows standard preliminary screening practice in structural equation modeling, in which zero-order correlations are examined before finalizing the structural model \citep{Tabachnick2018}. Because partial correlations are mathematically bounded by the zero-order correlation and typically shrink further once shared variance with other predictors is accounted for \citep{Bork2018}, a construct with a non-significant, near-zero zero-order correlation with Critical Thinking Enactment has little realistic prospect of yielding a meaningful path coefficient in the SEM. Constructs whose zero-order correlation with Critical Thinking Enactment did not reach statistical significance ($p < .05$) were therefore dropped before specifying the structural model at this stage. This step excludes Ability to Remember (TMS-GenAI) ($r = -.031$, ns), Collaboration (TMS-GenAI) ($r = .094$, ns) and Credibility (TMS-GenAI) ($r = .027$, ns). Six constructs were retained: GenAI Collaboration Confidence ($r = .080$, $p < .05$), GenAI Anchoring--CA-GenAI ($r = .098$, $p < .01$), Reflective Transformation--CA-GenAI ($r = .143$, $p < .01$), Ability to Think (TMS-GenAI) ($r = .099$, $p < .01$), Specialization \& Coordination (TMS-GenAI) ($r = .173$, $p < .01$), and GenAI Familiarity (TMS-GenAI) ($r = .119$, $p < .01$).

\subsubsection{Step 2---Same-Instrument Redundancy Check.} Among the six constructs retained in Step~1, we examined pairwise correlations specifically between constructs derived from a single item pool via exploratory factor analysis, since such pairs---whether representing convergent facets or opposing poles of one underlying tension---are prone to redundancy driven partly by shared method variance rather than distinct substantive content. This check identified one such pair: GenAI Anchoring and Reflective Transformation, both derived from the same six-item ``Cognitive Autonomy in GenAI-assisted Work (CA-GenAI)'' block, were highly correlated ($r = .623$) using the factor scores generated. We interpret this in light of the AI-autonomy literature, which distinguishes between AI's mediating influence on the internal, deliberative dimension of autonomy and the individual's exercise of independent judgment and control over the resulting output. This zero-order correlation between the factor scores ($r = .623$) is consistent in direction and magnitude with the rotated inter-factor correlation reported in Measures ($\varphi = .491$), both indicating a moderate-to-strong positive relationship between the two factors and supporting the convergent validity of the factor structure. The somewhat higher value obtained from the factor scores is expected given factor score indeterminacy, as regression-based scoring weights borrow shared variance from cross-loading items to maximize predictive validity, which can elevate the empirical correlation above the model-implied $\varphi$ \citep{Grice2001,Beauducel2026,Rhemtulla2025}. We retained the member of the pair with the stronger association with Critical Thinking Enactment---Reflective Transformation ($r = .143$) over GenAI Anchoring ($r = .098$)---and dropped GenAI Anchoring. Five constructs remained: GenAI Collaboration Confidence, Reflective Transformation (CA-GenAI), Ability to Think (TMS-GenAI), Specialization \& Coordination (TMS-GenAI), and GenAI Familiarity (TMS-GenAI). Although three of these five---Ability to Think, Specialization \& Coordination, and GenAI Familiarity---originate from the same 27-item TMS-GenAI pool, none showed the same-instrument redundancy pattern identified above for CA-GenAI. Thus, all five constructs were retained as the final construct set for structural modeling.

\subsubsection{Step 3 — Directional Structure via Established Accounts.}
With the final construct set fixed, we specified the directional relationships among them based on established process accounts in the relevant literature.

Classic TMS scholarship describes the underlying construct as a relational structure that emerges through accumulated interaction with a partner rather than a static individual trait \citep{Wegner1987}; subsequent work formalized this as manifesting in three interrelated dimensions---specialization, coordination, and credibility---each shown to develop through shared interaction history rather than existing independently of it \citep{Lewis2003}. A dyad's degree of specialization or coordination is, by definition, a property of its interaction history and cannot precede that history. We treat GenAI Familiarity as the operational proxy for this accumulated interaction history in a human-AI dyad, since it captures the extent to which a person has developed practical skill in retrieving and organizing information through GenAI. Following the TMS process sequence, Familiarity is modeled as prior to Specialization \& Coordination, rather than the reverse.

The same precedence logic extends to Reflective Transformation, though on different grounds than for Specialization \& Coordination. Reflective Transformation is not itself a TMS-derived construct, but its measurement structurally presupposes prior exposure to GenAI output: all three retained items ask respondents to characterize their reaction to an already-received AI-generated draft---considering alternatives beyond what the AI suggested, transforming the AI's suggestions, and being aware of the AI's influence on one's own originality. A respondent cannot meaningfully answer these items without having first encountered at least one instance of GenAI output to react to. Reflective Transformation is therefore modeled, like Specialization \& Coordination, as downstream of Familiarity---but as a parallel, independent elaboration of that experience, since neither construct is a precondition for the other.

Beyond this temporal precedence, Reflective Transformation's conceptual content also motivates its distinct positioning relative to Critical Thinking Enactment. Following \citet{Raz1986}'s self-authorship model---the fuller three-condition framework underlying the competency/independence distinction introduced in Section~2.4---autonomy can be decomposed into three conditions: rationality (understanding and acting on one's own reasons), optionality (the range of alternatives one considers), and independence (freedom from undue external influence). The retained Reflective Transformation items---considering approaches beyond the AI's suggestion, transforming that suggestion with one's own ideas, and maintaining awareness of AI's influence while producing original work---map most directly onto the rationality condition: the exercise of reasoned, deliberate judgment over an externally supplied input. Because Critical Thinking Enactment itself is fundamentally an exercise of this same reasoned, deliberate judgment, the two constructs share conceptual ground that a construct tapping mainly the optionality and independence conditions (such as the GenAI Anchoring construct dropped in Step~2) would not share to the same degree. Reflective Transformation is therefore modeled as a proximal antecedent of Critical Thinking Enactment: it remains downstream of Familiarity, but it sits conceptually and empirically closer to the outcome construct than the other retained predictors. This positioning is consistent with the broader AI-autonomy literature's distinction between AI's influence on deliberation itself and AI's influence on the practical exercise of judgment \citep{Catena2026}.

Ability to Think and GenAI Collaboration Confidence were positioned as antecedents to GenAI Familiarity, though the basis for this ordering differs between the two. Ability to Think reflects a domain-general self-appraisal of one's own cognitive competence that is conceptually prior to, and does not presuppose, any specific history of GenAI use. We nonetheless model Confidence as antecedent to Familiarity, because the three items composing this construct jointly describe a motivational sequence that arguably precedes accumulated experience with the tool. Confidence in one's own ability to complete a task without GenAI reasonably supports confidence in one's ability to evaluate GenAI's output, because a person who trusts their own independent judgment of the task is better positioned to assess whether GenAI's output meets that same standard. When this evaluative confidence is paired with confidence in GenAI's own capability to perform the task, the combination is likely to increase a person's willingness to engage with and rely on GenAI in the first place---they believe both that GenAI is capable and that they themselves are equipped to verify its output rather than accept it uncritically. It is this willingness to engage, prior to sustained use, that we take to be the proximal driver of the accumulated interaction captured by GenAI Familiarity, rather than the reverse. Although the three Confidence items load onto a single empirical factor (see Measures), we draw on their differentiated content only to motivate this temporal position relative to Familiarity, not to reintroduce three separable sub-constructs into the structural model. We treat this ordering as a testable modeling assumption rather than a settled causal claim.

\begin{figure*}
  \centering
  \includegraphics[width=\textwidth]{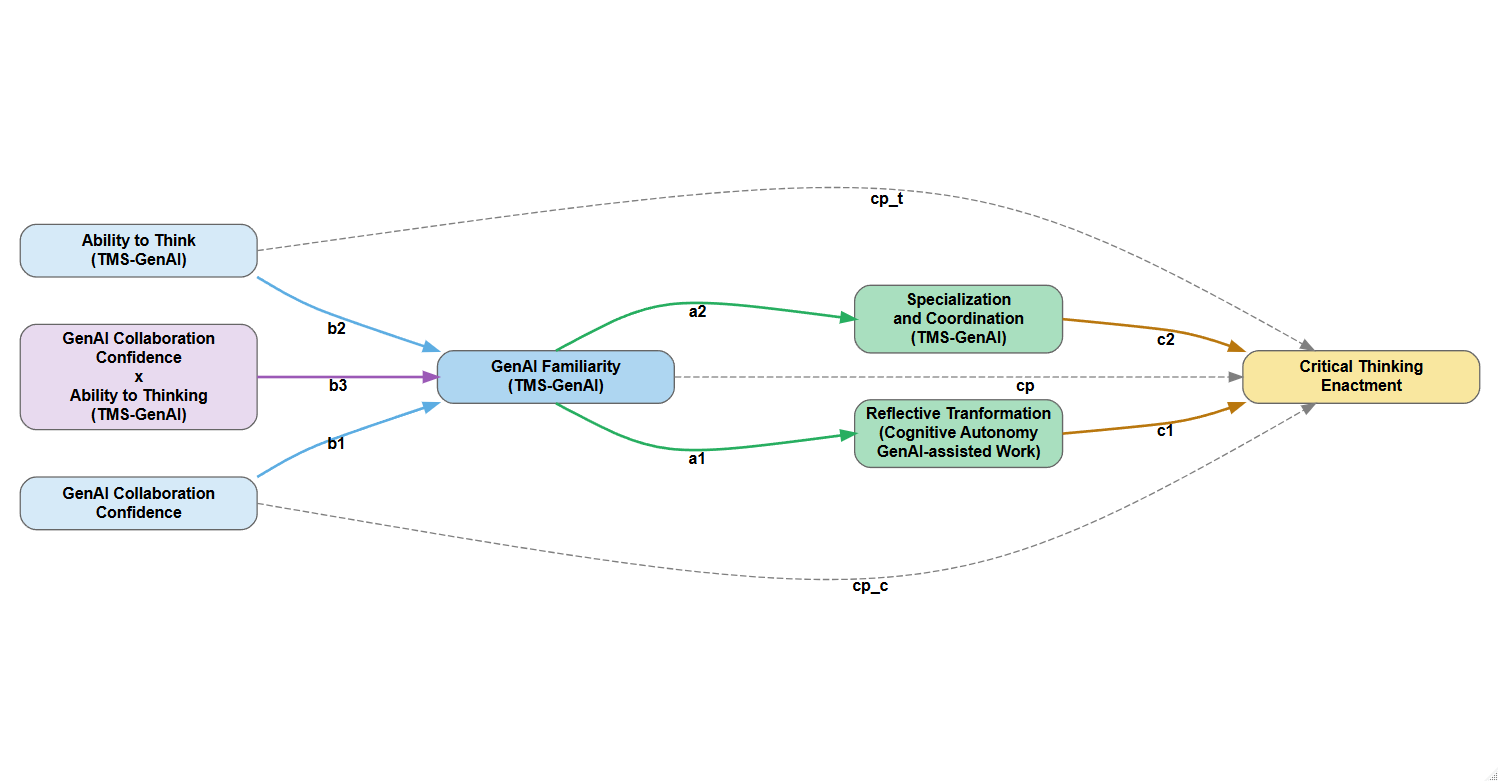}
  \caption{\textbf{Conceptual baseline model.} }
  \label{fig:conceptual-model}
\end{figure*}

The model also incorporates a moderation test and a parallel mediation test as two distinct analytic components. The interaction term between GenAI Collaboration Confidence and Ability to Think ($b_3$) tests whether Ability to Think moderates the strength of Confidence's effect on GenAI Familiarity, following the standard approach of entering a product term to capture the conditional effect of a moderator on a predictor--outcome path \citep{Aiken1991}. Separately, the paths linking Ability to Think, GenAI Collaboration Confidence, and GenAI Familiarity to Critical Thinking Enactment through the two parallel mediators, Specialization \& Coordination and Reflective Transformation, are evaluated as indirect effects using bootstrapped confidence intervals, consistent with recommended practice for testing multiple/parallel mediation without assuming normality in the sampling distribution of the indirect effect \citep{Preacher2008}. The overall model is illustrated in Figure~\ref{fig:conceptual-model}.

\subsection{Data Analysis}
The analysis proceeded in two stages. First, a baseline path model was estimated using the full pooled sample ($N = 891$) to examine the structural relationships among GenAI Collaboration Confidence, Ability to Think (TMS-GenAI), the interaction term of these two, GenAI Familiarity, Reflective Transformation (CA-GenAI), Specialization and Coordination (TMS-GenAI), and Critical Thinking Enactment. Second, a multigroup structural equation model (SEM) was estimated to assess whether these structural relations were equivalent across the United States and T{\"u}rkiye groups.

Given the binary nature of the Critical Thinking Enactment outcome, all models were estimated using the diagonally weighted least squares (DWLS) estimator with robust standard errors in \texttt{lavaan} 0.6-21, a choice consistent with simulation evidence showing that DWLS with robust corrections outperforms maximum likelihood when observed variables are categorical or binary \citep{Li2016}. This approach is also supported by comparative work demonstrating the relative advantages of categorical DWLS over robust maximum likelihood estimation under non-normal, ordinal data conditions \citep{Bandalos2014}, and by more recent guidance on correctly deriving robust standard errors for categorical-data SEM to avoid common estimation errors \citep{Lai2023}. The Delta parameterization was used, and model identification followed the standard variance-fixing convention for categorical outcomes, in line with recommended practice for mixed continuous-categorical SEM specifications \citep{Li2021}. The multigroup analysis was tested as a freely estimated model, a fully constrained model, and then a partial invariance model after path-by-path testing identified the non-invariant regression paths, following established procedures for evaluating measurement and structural invariance via chi-square-difference testing in multigroup SEM \citep{Yuan2016}.

Moreover, a moderation hypothesis was embedded in the structural model by including the multiplicative product term of GenAI Collaboration Confidence and Ability to Think (TMS-GenAI). This interaction term was created by first mean-centering both predictors, following \citet{Aiken1991}'s procedure for testing and interpreting interactions in multiple regression, a method also endorsed for its conceptual clarity by \citet{Sinacore1993} and \citet{Dalal2012}. It is important to note that mean-centering was used here strictly for interpretive purposes rather than as a remedy for multicollinearity. Methodological work has established that centering does not reduce essential collinearity between predictors and their product terms, since it leaves the determinant of the predictor cross-product matrix unchanged \citep{Echambadi2007,Iacobucci2016,Shieh2011}. Rather, centering was adopted because it renders the lower-order regression coefficients interpretable as conditional effects evaluated at the sample mean, which is a necessary condition for meaningful simple slope analysis \citep{Aiken2000}. Consistent with the caution raised by \citet{Wurm2025} that centered coefficients are sometimes misread as unconditional main effects, we interpret all lower-order coefficients strictly as effects conditional on the moderator's mean value, not as overall main effects. Simple slope analysis was then conducted at low ($-1$~SD) and high ($+1$~SD) values of the moderator to probe the conditional effect of GenAI Collaboration Confidence on GenAI Familiarity. In the multigroup analysis, group-specific slopes were computed separately for the United States and T{\"u}rkiye samples.

\section{Results}

\subsection{Baseline Model}
\subsubsection{Direct Effects}

The baseline structural model demonstrated excellent fit to the pooled sample ($\chi^2(8) = 24.258$, $p < .001$; CFI $= 0.974$; TLI $= 0.980$; RMSEA $= 0.048$; SRMR $= 0.024$), meeting conventional criteria for acceptable SEM fit. The structural paths are detailed in Table~\ref{tab:baseline-paths}.

\begin{figure}
  \centering
  \includegraphics[width=0.9\linewidth]{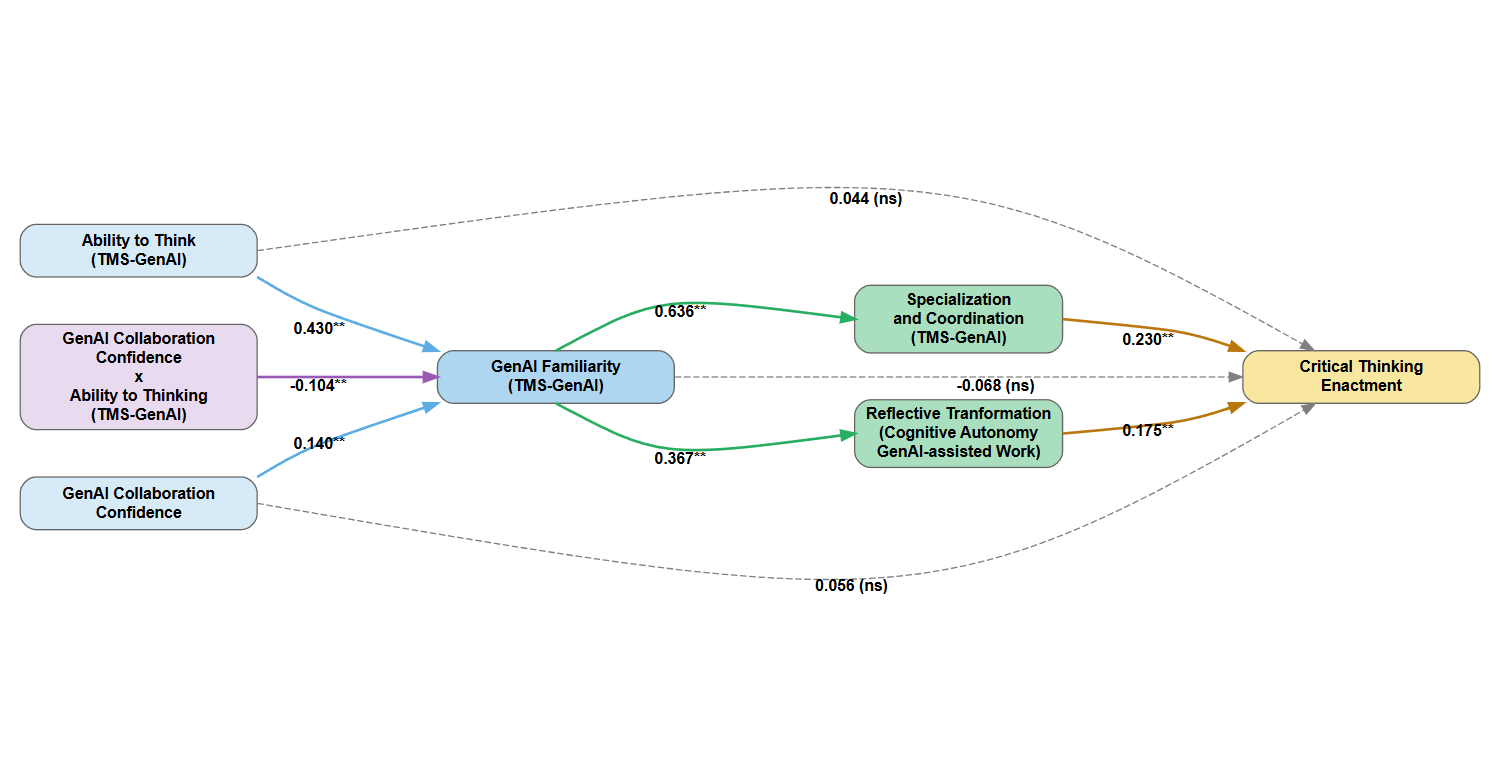}
  \caption{\textbf{Estimated baseline model.} }
  \label{fig:baseline-estimated}
\end{figure}

Ability to Think (TMS with GenAI) was the strongest predictor of GenAI Familiarity (TMS with GenAI), followed by GenAI Collaboration Confidence. The interaction term (Confidence $\times$ Thinking) exerted a significant negative effect on GenAI Familiarity, indicating a buffering moderation pattern. GenAI Familiarity, in turn, positively predicted both Specialization and Coordination (TMS with GenAI) and Reflective Transformation (CA-GenAI). Both mediators significantly predicted Critical Thinking Enactment, while the direct effects of GenAI Familiarity, GenAI Collaboration Confidence, and Ability to Think (TMS with GenAI) on Critical Thinking Enactment were non-significant, suggesting full mediation in the pooled model (see Figure~\ref{fig:baseline-estimated} for estimated model).

\begin{table}
\caption{\textbf{Baseline Model Regression Paths}}
\label{tab:baseline-paths}
\small
\begin{tabular}{@{}p{7.2cm} c p{3.4cm} rrrr@{}}
\toprule
\textbf{Path} & & \textbf{Outcome} & \textbf{Est.} & \textbf{SE} & \textbf{z} & \textbf{p} \\
\midrule
GenAI Collaboration Confidence ($b_1$) & $\rightarrow$ & GenAI Familiarity & 0.140 & 0.037 & 3.761 & .000 \\
Ability to Think (TMS-GenAI) ($b_2$) & $\rightarrow$ & & 0.430 & 0.030 & 14.445 & .000 \\
Interaction of Ability to Think (TMS-GenAI) and GenAI Collaboration Confidence ($b_3$) & $\rightarrow$ & & $-0.104$ & 0.027 & $-3.906$ & .000 \\
\midrule
GenAI Familiarity ($a_1$) & $\rightarrow$ & Reflective Transformation & 0.367 & 0.028 & 13.264 & .000 \\
GenAI Familiarity ($a_2$) & $\rightarrow$ & Specialization \& Coordination & 0.636 & 0.021 & 30.523 & .000 \\
\midrule
Reflective Transformation ($c_1$) & $\rightarrow$ & CT Enactment & 0.175 & 0.064 & 2.712 & .007 \\
Specialization and Coordination ($c_2$) & $\rightarrow$ & & 0.230 & 0.073 & 3.142 & .002 \\
GenAI Familiarity ($cp$) & $\rightarrow$ & & $-0.068$ & 0.089 & $-0.771$ & .440 \\
GenAI Collaboration Confidence ($cp_c$) & $\rightarrow$ & & 0.056 & 0.058 & 0.962 & .336 \\
Ability to Think (TMS-GenAI) ($cp_t$) & $\rightarrow$ & & 0.044 & 0.064 & 0.681 & .496 \\
\bottomrule
\end{tabular}
\end{table}

\subsubsection{Indirect Effects}
Mediation (see Table~\ref{tab:indirect-effects} for estimates) was tested using the product-of-coefficients approach with defined parameters, and confidence intervals were computed using the delta method. GenAI Familiarity significantly mediated the effect of Ability to Think (TMS) on Critical Thinking Enactment through both Reflective Transformation (CA-GenAI) and Specialization \& Coordination (TMS with GenAI). GenAI Familiarity itself also showed a significant total indirect effect on Critical Thinking Enactment through these same two mediators. The total indirect effect of Ability to Think (TMS with GenAI) on Critical Thinking Enactment was significant, and GenAI Collaboration Confidence also showed a significant total indirect effect via the same mediating structure.

\begin{table} [!htb]
\caption{\textbf{Baseline Model Indirect and Total Effects}}
\label{tab:indirect-effects}
\small
\begin{tabular}{@{}p{9.5cm} rrrr@{}}
\toprule
\textbf{Parameter} & \textbf{Est.} & \textbf{SE} & \textbf{z} & \textbf{p} \\
\specialrule{\heavyrulewidth}{0pt}{0pt}
GenAI Familiarity $\rightarrow$ Reflective Transformation $\rightarrow$ Critical Thinking Enactment & 0.064 & 0.025 & 2.616 & .009 \\
\cmidrule(lr){1-5}
GenAI Familiarity $\rightarrow$ Specialization \& Coordination $\rightarrow$ Critical Thinking Enactment & 0.147 & 0.047 & 3.149 & .002 \\
\cmidrule(lr){1-5}
\textit{Total Indirect (GenAI Familiarity)} & \textit{0.211} & \textit{0.055} & \textit{3.823} & \textit{.000} \\
\midrule
GenAI Collaboration Confidence $\rightarrow$ GenAI Familiarity $\rightarrow$ Reflective Transformation $\rightarrow$ Critical Thinking Enactment & 0.009 & 0.004 & 2.182 & .029 \\
\cmidrule(lr){1-5}
GenAI Collaboration Confidence $\rightarrow$ GenAI Familiarity $\rightarrow$ Specialization \& Coordination $\rightarrow$ Critical Thinking Enactment & 0.021 & 0.009 & 2.380 & .017 \\
\cmidrule(lr){1-5}
\textit{Total Indirect (GenAI Collaboration Confidence)} & \textit{0.029} & \textit{0.011} & \textit{2.667} & \textit{.008} \\
\midrule
Ability to Think (TMS-GenAI) $\rightarrow$ GenAI Familiarity $\rightarrow$ Reflective Transformation $\rightarrow$ Critical Thinking Enactment & 0.028 & 0.011 & 2.601 & .009 \\
\cmidrule(lr){1-5}
Ability to Think (TMS-GenAI) $\rightarrow$ GenAI Familiarity $\rightarrow$ Specialization \& Coordination $\rightarrow$ Critical Thinking Enactment & 0.063 & 0.020 & 3.076 & .002 \\
\cmidrule(lr){1-5}
\textit{Total Indirect (Ability to Think)} & \textit{0.091} & \textit{0.024} & \textit{3.719} & \textit{.000} \\
\bottomrule
\multicolumn{5}{@{}p{\dimexpr\textwidth-2\tabcolsep}@{}}{\footnotesize\textit{Note.} All indirect effects were estimated using the delta method.}
\end{tabular}
\end{table}

\subsubsection{Moderation – Simple Slope Analysis}
\begin{figure}[!htb]
  \centering
 \includegraphics[width=0.65\linewidth]{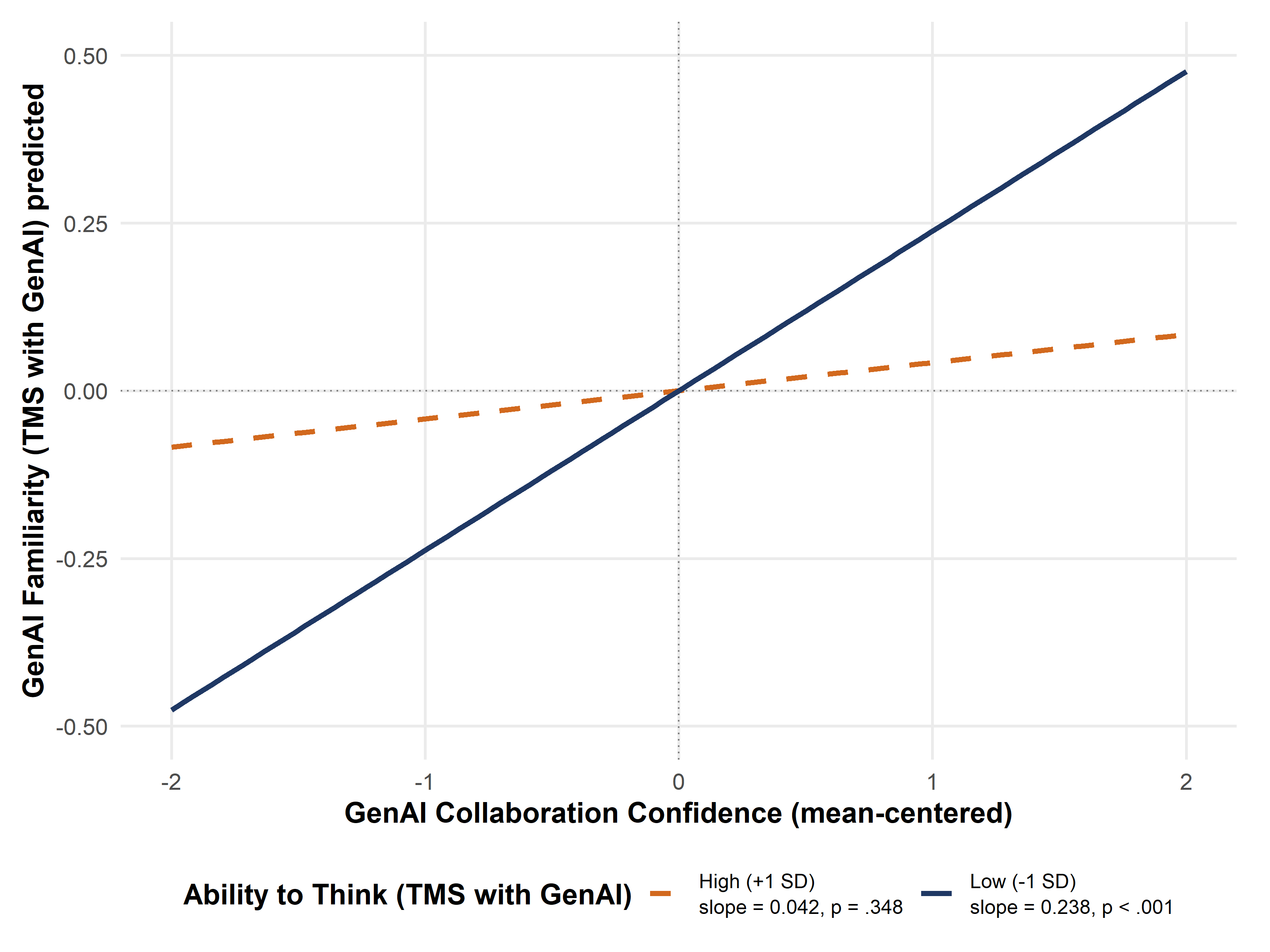}
  \caption{\textbf{Simple slope analysis.} }
  \label{fig:simpleslope-pooled}
\end{figure}

Simple slope analysis was conducted to test the GenAI Collaboration Confidence $\times$ Ability to Think (TMS with GenAI) interaction on GenAI Familiarity (TMS with GenAI) (see Figure~\ref{fig:simpleslope-pooled}). At low values of Ability to Think, the effect of GenAI Collaboration Confidence on GenAI Familiarity was significant and positive ($0.238$, $p < .001$), whereas at high values of Ability to Think the slope was reduced and non-significant ($0.042$, $p = 0.348$). This pattern is consistent with a buffering moderation effect in which the positive influence of the GenAI Collaboration Confidence on GenAI Familiarity weakens as Ability to Think (TMS with GenAI) increases.

\subsection{Multigroup Model: Structural Invariance}
The multigroup SEM analysis in Table~\ref{tab:fit-statistics} showed that the structural model was not fully invariant across the USA and T{\"u}rkiye groups. The fully constrained model fit values were worse than those of the freely estimated model, and releasing the two non-invariant paths produced a better-fitting partial invariance model. The partial model showed acceptable fit ($\chi^2(24) = 59.382$; CFI $= 0.938$; TLI $= 0.969$; RMSEA $= 0.058$; SRMR $= 0.035$).

\begin{table}[!htb]
\caption{\textbf{Model Fit Statistics for Baseline and Multigroup SEM}}
\label{tab:fit-statistics}
\small
\begin{tabular}{@{}l rrrr@{}}
\toprule
\textbf{Fit Index} & \textbf{Baseline Model} & \textbf{Free Multigroup Model} & \textbf{Partial Invariance Model} & \textbf{Fully Constrained Model} \\
\specialrule{\heavyrulewidth}{0pt}{0pt}
Model Parameters & 17 & 34 & 34 & 34 \\
Equality Constraints & 0 & 0 & 8 & 10 \\
$\chi^2$ (Standard) & 24.258 & 41.430 & 59.382 & 85.691 \\
$\chi^2$ (Scaled) & 37.877 & 56.256 & 64.792 & 86.894 \\
$df$ & 8 & 16 & 24 & 26 \\
CFI (Standard) & 0.974 & 0.955 & 0.938 & 0.895 \\
CFI (Scaled) & 0.941 & 0.916 & 0.915 & 0.874 \\
TLI (Standard) & 0.980 & 0.966 & 0.969 & 0.951 \\
TLI (Scaled) & 0.955 & 0.937 & 0.958 & 0.942 \\
RMSEA (Standard) & 0.048 & 0.060 & 0.058 & 0.072 \\
RMSEA (Scaled) & 0.065 & 0.075 & 0.062 & 0.073 \\
SRMR & 0.024 & 0.026 & 0.035 & 0.042 \\
\bottomrule
\multicolumn{5}{@{}p{\dimexpr\textwidth-2\tabcolsep}@{}}{\footnotesize\textit{Note.} Estimators for all models are DWLS. $N = 891$, USA $n = 645$, T{\"u}rkiye $n = 246$. CFI = Comparative Fit Index; TLI = Tucker--Lewis Index; RMSEA = Root Mean Square Error of Approximation; SRMR = Standardized Root Mean Square Residual. CFI and TLI $\geq 0.90$, RMSEA $\leq 0.08$, and SRMR $\leq 0.08$ indicate acceptable model fit.}
\end{tabular}
\end{table}

\begin{figure}[!htb]
  \centering
  \includegraphics[width=\linewidth]{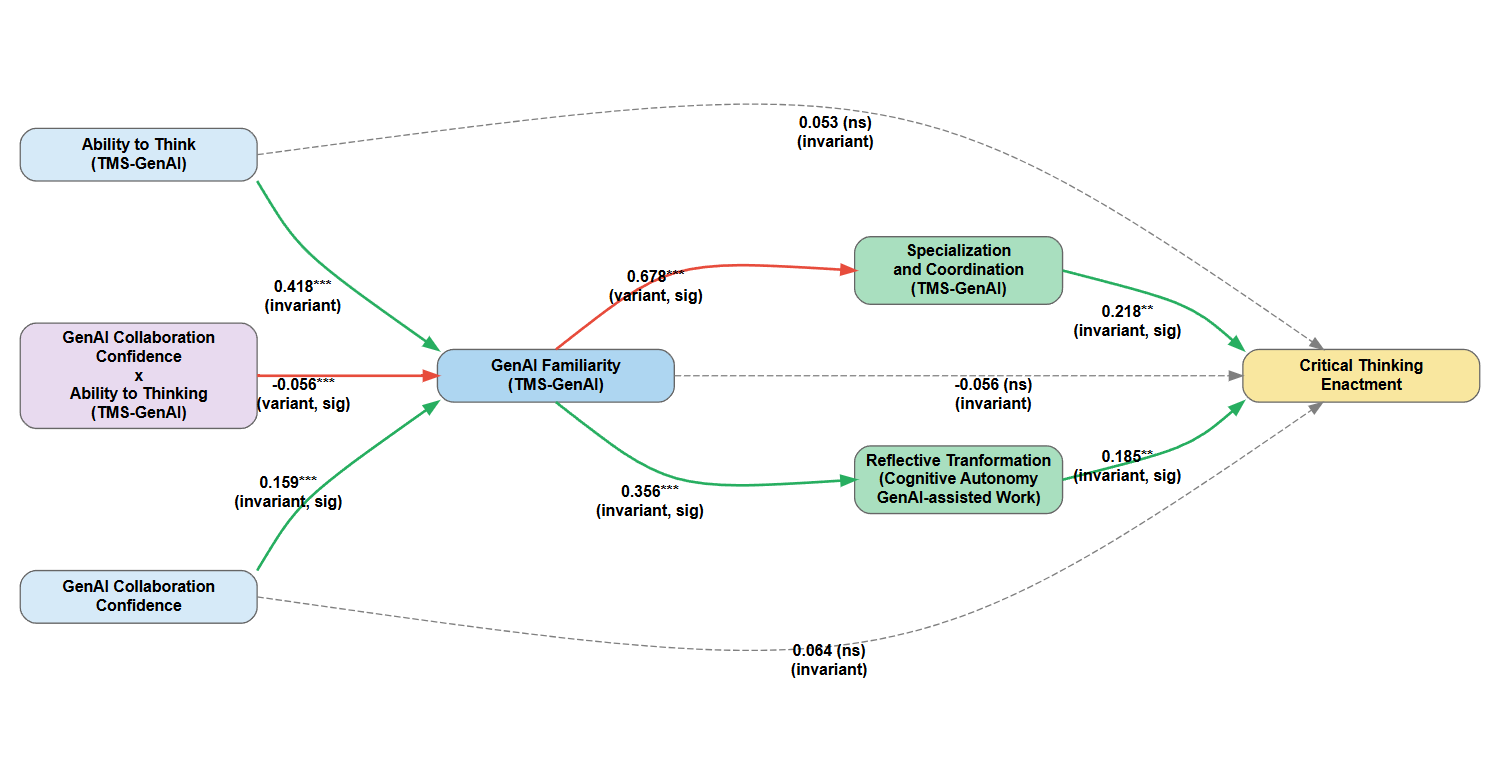}
  \caption{\textbf{The USA estimated model.}}
  \label{fig:usa-model}
\end{figure}

\begin{figure}[!htb]
  \centering
  \includegraphics[width=\linewidth]{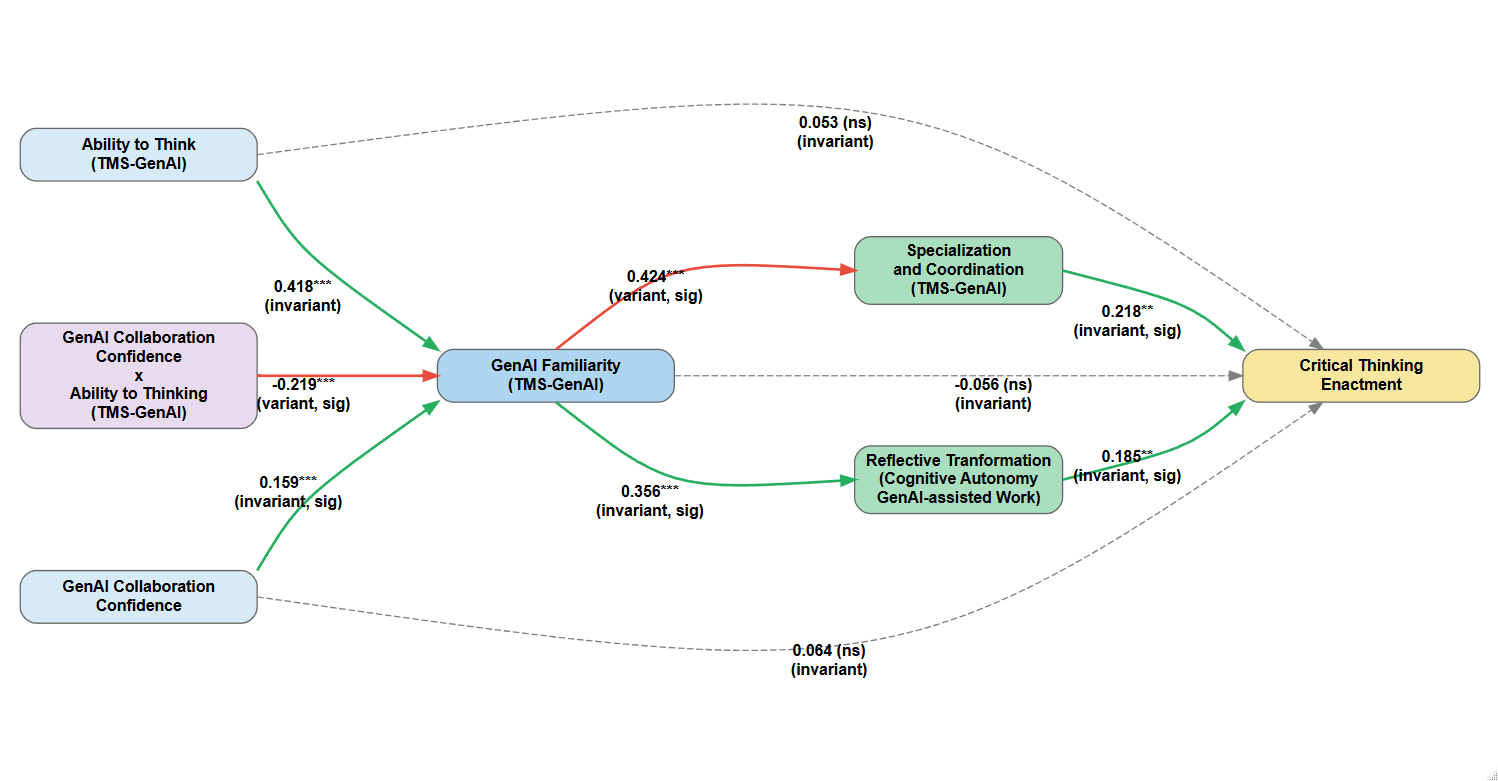}
  \caption{\textbf{The T\"urkiye estimated model.} }
  \label{fig:turkiye-model}
\end{figure}

As illustrated in Table~\ref{tab:multigroup-paths} and Table~\ref{tab:invariance-tests}, path-by-path tests further indicated that two regression paths were variant across groups: the interaction between GenAI Collaboration Confidence and Ability to Think (TMS with GenAI) on GenAI Familiarity (TMS with GenAI), $\chi^2(1) = 8.503$, $p = .004$, and GenAI Familiarity (TMS with GenAI) on Specialization \& Coordination (TMS with GenAI), $\chi^2(1) = 27.534$, $p < .001$. All other structural paths were invariant across the USA and T{\"u}rkiye groups. Moreover, the path-by-path tests indicated that the interaction between GenAI Collaboration Confidence and Ability to Think (TMS with GenAI) on GenAI Familiarity (TMS with GenAI) differed significantly between the two countries and had a more negative estimate in T{\"u}rkiye ($-0.219$) than in the USA ($-0.056$). GenAI Familiarity (TMS with GenAI) on Specialization \& Coordination (TMS with GenAI) also differed significantly, with a stronger positive effect in the USA ($0.678$) than in T{\"u}rkiye ($0.424$). The estimated models for both countries are illustrated in Figure~\ref{fig:usa-model} and Figure~\ref{fig:turkiye-model}.

\begin{table}[!htb]
\caption{\textbf{Multigroup Constrained Regression Paths (USA and T\"urkiye)}}
\label{tab:multigroup-paths}
\small
\begin{tabular}{@{}p{6.5cm} rrl rrl@{}}
\toprule
\multirow{2}{*}{\textbf{Path (Label)}} & \multicolumn{3}{c}{\textbf{USA}} & \multicolumn{3}{c}{\textbf{T\"urkiye}} \\
\cmidrule(lr){2-4} \cmidrule(lr){5-7}
 & \textbf{$\beta$} & \textbf{SE} & \textbf{p} & \textbf{$\beta$} & \textbf{SE} & \textbf{p} \\
\specialrule{\heavyrulewidth}{0pt}{0pt}
GenAI Collaboration Confidence $\rightarrow$ GenAI Familiarity ($b_1$) & 0.159 & 0.037 & $<.001$ & 0.159 & 0.037 & $<.001$ \\
Ability to Think (TMS-GenAI) $\rightarrow$ GenAI Familiarity ($b_2$) & 0.418 & 0.029 & $<.001$ & 0.418 & 0.029 & $<.001$ \\
Interaction $\rightarrow$ GenAI Familiarity ($b_3$) & $-0.056$ & 0.028 & .048 & $-0.219$ & 0.048 & $<.001$ \\
GenAI Familiarity $\rightarrow$ Reflective Transformation ($a_1$) & 0.356 & 0.028 & $<.001$ & 0.356 & 0.028 & $<.001$ \\
GenAI Familiarity $\rightarrow$ Specialization \& Coordination ($a_2$) & 0.678 & 0.026 & $<.001$ & 0.424 & 0.039 & $<.001$ \\
Reflective Transformation $\rightarrow$ Critical Thinking Enactment ($c_1$) & 0.185 & 0.065 & .004 & 0.185 & 0.065 & .004 \\
Specialization \& Coordination $\rightarrow$ Critical Thinking Enactment ($c_2$) & 0.218 & 0.077 & .005 & 0.218 & 0.077 & .005 \\
GenAI Familiarity $\rightarrow$ Critical Thinking Enactment ($cp$) & $-0.056$ & 0.096 & .558 & $-0.056$ & 0.096 & .558 \\
GenAI Collaboration Confidence $\rightarrow$ Critical Thinking Enactment ($cp\_c$) & 0.064 & 0.059 & .273 & 0.064 & 0.059 & .273 \\
Ability to Think (TMS-GenAI) $\rightarrow$ Critical Thinking Enactment ($cp\_t$) & 0.053 & 0.067 & .425 & 0.053 & 0.067 & .425 \\
\bottomrule
\multicolumn{7}{@{}p{\dimexpr\textwidth-2\tabcolsep}@{}}{\footnotesize\textit{Note.} The paths Interaction $\rightarrow$ GenAI Familiarity and GenAI Familiarity $\rightarrow$ Specialization \& Coordination were freed across groups because they were non-invariant.}
\end{tabular}
\end{table}

\begin{table}[!htb]
\caption{\textbf{Path-by-Path Invariance Tests for USA vs.\ T\"urkiye}}
\label{tab:invariance-tests}
\small
\begin{tabular}{@{}p{7cm} rrrl@{}}
\toprule
\textbf{Path} & \textbf{$\chi^2$ diff} & \textbf{$df$ diff} & \textbf{p} & \textbf{Invariant?} \\
\specialrule{\heavyrulewidth}{0pt}{0pt}
\textit{GenAI Familiarity $\rightarrow$ Specialization \& Coordination} & \textit{27.534} & \textit{1} & \textit{$<.001$} & \textit{No} \\
\textit{Interaction $\rightarrow$ GenAI Familiarity} & \textit{8.503} & \textit{1} & \textit{.004} & \textit{No} \\
Ability to Think (TMS-GenAI) $\rightarrow$ GenAI Familiarity & 3.767 & 1 & .052 & Yes \\
GenAI Collaboration Confidence $\rightarrow$ GenAI Familiarity & 3.049 & 1 & .081 & Yes \\
Ability to Think (TMS-GenAI) $\rightarrow$ Critical Thinking Enactment & 2.577 & 1 & .108 & Yes \\
Reflective Transformation $\rightarrow$ Critical Thinking Enactment & 2.351 & 1 & .125 & Yes \\
GenAI Familiarity $\rightarrow$ Reflective Transformation & 1.182 & 1 & .277 & Yes \\
GenAI Familiarity $\rightarrow$ Critical Thinking Enactment & 0.570 & 1 & .450 & Yes \\
GenAI Collaboration Confidence $\rightarrow$ Critical Thinking Enactment & 0.399 & 1 & .528 & Yes \\
Specialization \& Coordination $\rightarrow$ Critical Thinking Enactment & 0.256 & 1 & .613 & Yes \\
\bottomrule
\end{tabular}
\end{table}

\subsubsection{Group-Specific Simple Slopes}
\begin{figure}[!htb]
  \centering
  \includegraphics[width=\linewidth]{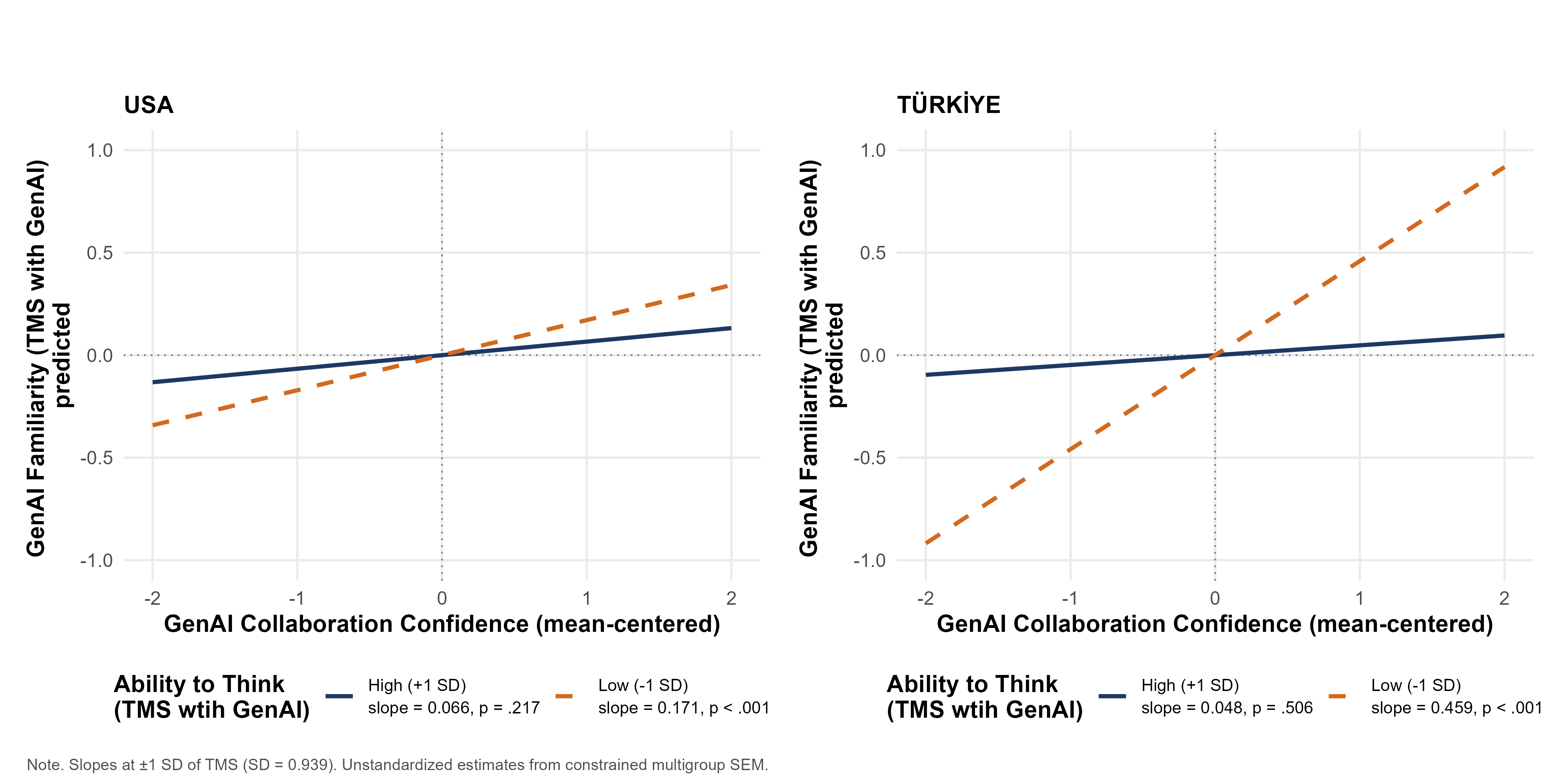}
  \caption{\textbf{Group-specific simple slopes.}}
  \label{fig:simpleslope-bygroup}
\end{figure}

Because of significant and variant results, group-specific simple slope analyses were used to probe the impact of the interaction between GenAI Collaboration Confidence and Ability to Think on GenAI Familiarity in each country (see Table~\ref{tab:group-slopes} and Figure~\ref{fig:simpleslope-bygroup}). In both countries, the slope for GenAI Collaboration Confidence was stronger when Ability to Think was low, and weaker when Ability to Think was high, but the pattern was more pronounced in T{\"u}rkiye. In the USA group, the low-Thinking slope was significant and positive, whereas the high-Thinking slope was not significant. In T{\"u}rkiye, the low-Thinking slope was substantially larger and highly significant, while the high-Thinking slope remained non-significant.

\begin{table}[!htb]
\centering
\begin{threeparttable}
\caption{\textbf{Group-Specific Simple Slope Analysis for the Moderation of Ability to Think on the Association between GenAI Collaboration Confidence and GenAI Familiarity in USA and T\"urkiye}}
\label{tab:group-slopes}
\small
\begin{tabular}{@{}l rrrr rr@{}}
\toprule
\textbf{Label} & \textbf{$\beta$} & \textbf{SE} & \textbf{z} & \textbf{p} & \textbf{CI Lower} & \textbf{CI Upper} \\
\specialrule{\heavyrulewidth}{0pt}{0pt}
Low Ability to Think in USA & 0.171 & 0.050 & 3.433 & .001 & 0.073 & 0.268 \\
High Ability to Think in USA & 0.066 & 0.053 & 1.235 & .217 & $-0.039$ & 0.170 \\
Low Ability to Think in T\"urkiye & 0.459 & 0.082 & 5.566 & .000 & 0.297 & 0.621 \\
High Ability to Think in T\"urkiye & 0.048 & 0.072 & 0.665 & .506 & $-0.093$ & 0.189 \\
\bottomrule
\end{tabular}
\begin{tablenotes}
\small
\item \textit{Note.} Low and high moderator values correspond to $\pm 1$~SD of Ability to Think (TMS with GenAI); SD $= 0.939$.
\end{tablenotes}
\end{threeparttable}
\end{table}

\section{Discussion}
\subsection{Antecedents of GenAI Familiarity: Ability to Think and Confidence}

Ability to Think (TMS-GenAI) emerged as the strongest predictor of GenAI Familiarity in both the pooled and multigroup analyses, consistent with Transactive Memory Systems theory's premise that individuals who distribute cognitive labor across a network of ``who and what knows'' are better positioned to recognize and draw on external sources of expertise, human or technological \citep{Wegner1987,Ren2011}. This aligns with a companion validation study introducing the TMS-GenAI measure itself, which found thinking ability to be a stable construct across the Turkish and U.S. samples even as more collaboration-specific dimensions varied culturally \citep{Akgun2026}.

The significant negative interaction between GenAI Collaboration Confidence and Ability to Think indicates a buffering effect. Confidence matters more for building GenAI Familiarity when a person's self-appraised thinking ability is weaker, and matters less once that thinking-ability orientation is already strong. Substantively, confidence appears to let users with a weaker distributed-thinking disposition ``act their way into'' GenAI familiarity, consistent with self-efficacy functioning as a compensatory rather than additive resource in technology adoption \citep{Wang2023,Akben2025}. This buffering pattern was significantly stronger in T{\"u}rkiye than in the United States; we return to that cross-national difference below, where it belongs alongside the study's other invariance finding.

\subsection{Mediated Pathways to Critical Thinking Enactment}

The direct effects of GenAI Familiarity, GenAI Collaboration Confidence, and Ability to Think on Critical Thinking Enactment were not significant in the pooled model, whereas their indirect effects through Specialization \& Coordination and Reflective Transformation were. This pattern indicates full mediation. Familiarity, confidence, and thinking ability shape critical thinking enactment only through how a person structures their engagement with GenAI, not directly. The stronger mediated route ran through Specialization \& Coordination, suggesting GenAI Familiarity becomes consequential for critical thinking specifically when it helps users divide cognitive labor and structure their interaction with GenAI more deliberately. Reflective Transformation contributed a second, parallel route through active reconsideration of the AI's output.

This is consistent with a distributed-cognition view in which critical engagement emerges from how tasks and representations move across user and AI, not from either agent's isolated capability \citep{Zhao2026}, and it carries a cautionary edge. \citet{Gerlich2025} found frequent AI use predicts lower critical thinking via cognitive offloading, and the Specialization \& Coordination pathway identified here may represent the protective counter-mechanism. Users who deliberately structure their division of labor with GenAI may avoid the offloading trajectory Gerlich describes, whereas those who fail to do so remain at risk of it. \citet{Kim2025}'s deskilling/upskilling framework captures the same point. GenAI familiarity alone is valence-neutral, and whether it produces dependence-driven deskilling or capability-building upskilling depends on the structure of use, not on familiarity's magnitude. Together, this reframes the study's central contribution. Critical thinking enactment in GenAI-assisted work is a function of how deliberately users organize their cognitive partnership with GenAI, not of how much they know about it or how confident they feel using it.

\subsection{Cross-National Invariance}

The multigroup model was partially invariant. Most structural paths held equally across the United States and T{\"u}rkiye, but two did not---the Confidence $\times$ Ability to Think interaction on GenAI Familiarity, and the path from GenAI Familiarity to Specialization \& Coordination. Partial invariance of this kind is the norm rather than the exception in cross-national SEM, and established guidance treats a small number of theoretically interpretable non-invariant paths as an acceptable basis for group comparison rather than a modeling failure \citep{Henseler2016,Pokropek2019}.

The confidence-buffering effect was significantly stronger in T{\"u}rkiye than in the United States. One possible cultural account is that individualist contexts may lead users to construe AI as external to the self, while more collectivist contexts may fold AI more readily into a socially embedded way of working, making confidence a more decisive lever where distributed-thinking orientation is weaker \citep{Barnes2024}. The familiarity-to-specialization-and-coordination path was stronger in the United States ($0.678$) than in T{\"u}rkiye ($0.424$); \citet{Jarvenpaa2011}'s case study of a culturally mixed engineering team shows the same mechanism concretely---in higher-power-distance, more collectivist settings, specialization and coordination may be partly governed by pre-existing social structure rather than driven purely by individual familiarity. Both deviations therefore look like theoretically interpretable cultural boundary conditions rather than measurement artifacts, consistent with the broader expectation that cross-national comparisons pairing a Western and a non-Western context typically surface partial, not full, invariance \citep{Shela2023}.

\subsection{Limitations and Future Directions}

Several limitations qualify these findings. The cross-sectional design precludes causal inference, and the binary operationalization of Critical Thinking Enactment limits measurement precision relative to a continuous scale. The Turkish sample was smaller than the the United States sample, which may have reduced power for detecting group-specific effects and warrants caution when interpreting the strength, though not the direction, of the country-specific slopes. As is often the case for initial studies introducing a new cross-cultural framework, the cultural explanations offered here for the two non-invariant paths draw on adjacent literatures rather than direct tests of the TMS-GenAI framework itself, and future work applying TMS-GenAI across additional cultural contexts would help establish these mechanisms more directly. The confidence-buffering effect, in particular, is read here only in terms of its benefit for building GenAI Familiarity; a broader literature on AI self-efficacy shows the same compensatory belief can carry different, even counterproductive, consequences for outcomes such as job insecurity or dependency \citep{Liu2026,Zhang2024}, so its benefits should not be assumed to extend automatically to well-being or autonomy outcomes this study did not measure. Future research should address these limitations in three ways. First, replicating the model longitudinally would test whether the confidence-buffering effect and its cross-national asymmetry are stable over time. Second, extending the confidence-buffering effect's outcome set beyond GenAI Familiarity to well-being and Critical Thinking Enactment would test this boundary condition directly. Third, collecting matched-size cross-national samples would help isolate cultural effects from sampling artifacts.

\subsection{Implications}

\textbf{Theoretical implications.} This study extends Transactive Memory Systems theory beyond human-human collaboration into human-GenAI partnership, showing that self-appraised TMS-based thinking, not mere familiarity or confidence, is the strongest driver of how users engage GenAI as a knowledge partner. It also clarifies the mechanism by which GenAI familiarity translates into critical thinking. The fully mediated pathway through Specialization \& Coordination and Reflective Transformation indicates that critical thinking enactment is a function of how users structure their cognitive partnership with GenAI, not how much they know about it. The confidence-TMS interaction further nuances self-efficacy theory by showing confidence operates as a compensatory, not additive, resource. Finally, the partial cross-national invariance demonstrates that TMS-GenAI mechanisms are largely culturally robust, while specific conversion pathways are culturally conditioned in ways consistent with power-distance and individualism-collectivism theory.

\textbf{Practical implications.} For higher-education instructors and curriculum designers, the findings suggest GenAI literacy interventions should prioritize deliberate task-structuring and distributed-thinking skills, not just platform familiarity or user confidence, since familiarity and confidence alone did not translate into critical thinking gains here. For institutions operating across cultural contexts, the confidence-buffering effect suggests confidence-building interventions may hold particular value for learners with weaker self-appraised thinking orientation, especially in higher-collectivism, higher-power-distance settings such as T{\"u}rkiye. 

\bibliographystyle{ACM-Reference-Format}
\bibliography{references}
\appendix

\section{Survey Instrument}
\label{app:instrument}

\subsection*{GenAI Collaboration Confidence}
\textit{Scale: Not at all confident (1), Slightly confident (2), Moderately confident (3), Very confident (4), Extremely confident (5)}
\begin{enumerate}
\item How confident are you in your ability to complete this academic task without GenAI?
\item How confident are you in the ability of GenAI to complete this academic task?
\item How confident are you in your ability to evaluate the accuracy and quality of the output that GenAI produces for this academic task?
\end{enumerate}

\subsection*{Cognitive Autonomy in GenAI-assisted Work (CA-GenAI)}
\textit{Scale: Not at all (1), To a small extent (2), To a moderate extent (3), To a considerable extent (4), To a great extent (5)}

\textit{\textbf{GenAI Anchoring}}
\begin{enumerate}
\item The AI-generated first draft tends to shape the overall direction of my final work
\item I find myself mostly refining the AI's approach rather than exploring fundamentally different alternatives
\item The AI's initial suggestions limit my creativity or originality
\end{enumerate}

\textit{\textbf{Reflective Transformation}}
\begin{enumerate}
\item I often consider multiple different approaches beyond what the AI initially suggests
\item The AI's suggestions serve as a starting point that I significantly transform with my own ideas
\item I'm aware of being influenced by the AI's suggestions even when trying to create something original
\end{enumerate}

\subsection*{Transactive Memory Systems with GenAI (TMS-GenAI)}
\textit{Scale: Strongly disagree (1), Disagree (2), Neutral (3), Agree (4), Strongly agree (5)}

\textit{\textbf{Ability to Think}}
\begin{enumerate}
\item I am smart.
\item I am smarter than the average person.
\item My mind is one of my best qualities.
\item I am good at thinking.
\item I feel good about my ability to think through problems.
\item I am capable of solving most problems without outside help.
\end{enumerate}

\textit{\textbf{Ability to Remember/Memory}}
\begin{enumerate}
\item I am proud of my memory.
\item I feel good about my ability to remember things.
\item I have a better memory than most people.
\item I have a good memory for recalling trivial information.
\end{enumerate}

\textit{\textbf{GenAI Familiarity}}
\begin{enumerate}
\item I know how to use Generative AI to answer questions I don't immediately know.
\item When I don't know the answer to a question right away, I know how to find it using Generative AI.
\item I know which Generative AI tools to use when I need to retrieve information efficiently.
\item I am skilled at using Generative AI to track down and organize information.
\end{enumerate}

\textit{\textbf{Credibility}}
\begin{enumerate}
\item I trust that the information provided by the Generative AI tool is credible.
\item I am confident in relying on the Generative AI tool's knowledge when learning new concepts.
\item I have a lot of faith in the Generative AI tool's ability to provide accurate and useful knowledge.
\end{enumerate}

\textit{\textbf{Collaboration}}
\begin{enumerate}
\item There are very few misunderstandings when I interact with the Generative AI tool for learning.
\item The Generative AI tool and I usually complete learning tasks efficiently.
\item I can seamlessly integrate the AI's responses into my learning process.
\item There is little confusion when I use the Generative AI tool to assist me in learning new topics.
\end{enumerate}

\textit{\textbf{Specialization and Coordination}}
\begin{enumerate}
\item The Generative AI tool and I have specialized knowledge in different domains.
\item I rely on the Generative AI tool for information and expertise that I do not personally possess.
\item The Generative AI tool and I are responsible for expertise in different aspects of the learning process.
\item To complete learning tasks, both my knowledge and the Generative AI tool's specialized knowledge are needed.
\item I know which topics or tasks I should rely on the Generative AI tool for assistance.
\item The Generative AI tool and I work together in a well-coordinated manner to complete learning tasks.
\end{enumerate}
\end{document}